\documentclass[aps,prd,twocolumn,amsmath,amssymb,showpacs]{revtex4}
\usepackage{graphicx,slashed,cancel,bbm}
\newcommand{\be}{\begin{equation}}
\newcommand{\ee}{\end{equation}}
\newcommand{\bea}{\begin{eqnarray}}
\newcommand{\eea}{\end{eqnarray}}

\newcommand{\sfrac}[2]{{\textstyle\frac{#1}{#2}}}
\newcommand{\D}{\mathrm{d}}
\newcommand{\E}{\mathrm{e}}
\newcommand{\I}{\mathrm{i}}

\newcommand{\Lag}{\mathcal{L}}

\begin{document}
\title{Worldline holographic Schwinger effect}
\author{Dennis D.~Dietrich}
\affiliation{Arnold Sommerfeld Center, Ludwig-Maximilians-Universit\"at, M\"unchen, Germany}
\affiliation{Institut f\"ur Theoretische Physik, Goethe-Universit\"at, Frankfurt am Main, Germany}
\begin{abstract}
 The decay of the vacuum due to the presence of an electric field is expected to be delayed by a confining force. We demonstrate that this feature is captured by our model \cite{Dietrich:2014...} for hadrons based on the worldline formalism. Our model, while based entirely in four-dimensional quantum field theory, shares many features with holographic approaches: it appears intrinsically quantum mechanical; as an auxiliary fifth dimension Schwinger's proper time combines with the physical four spacetime dimensions into an AdS$_5$ geometry; conformal-symmetry breaking contributions lead to warping; hidden local symmetry emerges; four-dimensional sources are extended to five-dimensional fields by a Wilson flow (gradient flow); and a variational principle for this flow reproduces the corresponding holographic calculation.
The approach also yields the higher-dimensional description in the nonrelativistic case.
\end{abstract}
\maketitle
\allowdisplaybreaks

\section{Introduction}

The instability of the vacuum against the production of particles in the presence of certain external fields, the Euler-Heisenberg-Schwinger effect \cite{Sauter:1931zz,Dunne:2004nc}, was one of the first nontrivial predictions of quantum field theory. 
Despite its long history, the effect still awaits experimental verification, e.g., with the help of ultrastrong light sources \cite{ELI}. 
The physical picture behind the effect is that in the presence of an electric field a virtual particle-antiparticle pair can gain the necessary amount of energy, $\propto m$, over one Compton wavelength, $\propto m^{-1}$, and thus become a real (on-shell) pair. Being a tunneling effect, the process should be suppressed exponentially with an exponent proportional to the ratio of the energy required (to put the pair on the mass shell) and the energy gained (over one Compton wavelength in the field). For a constant electric field this is proportional to $\propto m/(E\times\frac{1}{m})=\frac{m^2}{E}$, which is confirmed by detailed computations.

In the context of quantum chromodynamics the effect is of interest, for instance, in ultrarelativistic collisions. In quantum chromodynamics for covariantly constant field configurations, the standard computation proceeds in close analogy to the static Abelian case \cite{nabeffact}. What must, however, still be taken into account is the presence of a confining interaction, which forestalls the production of free quark-antiquark pairs. This is largely independent of whether the destabilising field itself is electromagnetic or chromoelectromagnetic. The reason for the delay is that when the virtual pair is being separated the energy gained in the external field is reduced by the energy lost by working against the confining force. {The vacuum persistence has been revisited in holographic settings \cite{Semenoff:2011ng,Gorsky:2001up,Sato:2013dwa,Kawai:2013xya,Hashimoto:2013mua} and a delay of vacuum decay to larger field strengths in the presence of a confining force has been found \cite{Sato:2013dwa,Kawai:2013xya,Hashimoto:2013mua}.
(This delayed onset of the vacuum decay is distinct from another threshold, first diagnosed in a holographic context in \cite{Semenoff:2011ng}, above which the decay of the Schwinger effect is found to change qualitatively.)}

Therefore, we analyse here what this means for our model of hadrons \cite{Dietrich:2014...}, which we derived in the worldline formalism of quantum field theory. Although our model is derived entirely in four-dimensional quantum field theory without recourse to input from gauge/gravity or gauge/string dualities, it shares many features with those holographic approaches. We summarise briefly the ideas behind the model in Sect.~\ref{sec:nostrings}. Indeed our model predicts the delay to stronger fields of the destabilisation of the vacuum. 

Before the introduction to our model in Sect.~\ref{sec:nostrings}, we rederive the standard Euler-Heisenberg-Schwinger result in Sect.~\ref{sec:persist}. In Sect.~\ref{sec:wh}, we demonstrate that our model shows confinement insofar as it delays particle production to higher field strengths. Sect.~\ref{sec:concl} concludes the paper.

\section{Vacuum persistence\label{sec:persist}}

Let us start by recapitulating how the original Euler-Heisenberg-Schwinger result \cite{Sauter:1931zz} comes about in detail. To this end we employ the worldline formulation \cite{Strassler:1992zr} of quantum field theory. There, in the presence of a gauge field $A^\mu(x)$ the Euclidean one-loop effective action reads
\begin{align}
w
=
&\int_{\varepsilon>0}^\infty\frac{\D T}{T^3}\;\E^{-m^2T}\int \D^4x_0\;\Lag,
\label{eq:ads?}
\\
\Lag
=
\frac{\mathcal{N}}{(4\pi)^2}
&\int_\mathrm{P}[\D y]\;\E^{-\int_0^T\D\tau[\frac{\dot y^2}{4}+\I \dot y \cdot A(x_0+y)]} ,
\label{eq:lag}
\end{align}
where we have used scalar matter for the sake of simplicity. The result does not change fundamentally for fermionic matter. Here $T$ is Schwinger's proper time. The integration is proper-time regularised by imposing $T\ge\epsilon>0$. $m$ stands for the mass of the elementary matter. The Lagrangian $\mathcal{L}$ consists of a path integral over all periodic particle trajectories. The normalisation $\mathcal{N}$ cancels the path integral for $A^\mu(x)\equiv 0$. The position $x^\mu$ is decomposed into $x_0^\mu+y^\mu$, where $x_0^\mu$ is the `centre-of-mass' coordinate, 
\be
\int_0^T\D\tau\,y^\mu=0 ,
\label{eq:com}
\ee
to remove the zero mode of the kinetic operator and make translational invariance as well as momentum conservation manifest. 

In a constant electric field $E$ in the direction of $x^1$ the interaction term can be written as
\be
2\I \dot y \cdot A(x_0+y)
=
E\,(\dot y^0y^1-y^0\dot y^1) .
\ee
The extra factor of $\I$ arises from the Wick rotation. The periodic orbits can be parametrised by
\be
y^\mu
=
\sum_{n=-\infty}^{+\infty}a_n^\mu\;\E^{\I n\frac{2\pi}{T}\tau} ,
\label{eq:Fourier}
\ee
where ${a_n^\mu}^*=a_{-n}^\mu$ to ensure that $y^\mu$ is real and $a_0^\mu\equiv 0$ because of (\ref{eq:com}). Accordingly, the path-integral measure becomes
\be
\int_\mathrm{P}[\D y]
=
\prod_{n=1}^\infty\int \D a_n^\mu\: \D (a_n^\mu)^*
=
\int [\D a] [\D a^*] .
\label{eq:pint}
\ee
With this parametrisation the worldline action in the exponent of the Lagrangian reads
\begin{widetext}
\begin{align}\nonumber
&-\frac{1}{4}\int_0^T\D\tau\;\Big[\dot y^2-2E(\dot y^0y^1-y^0\dot y^1)\Big]
=\\={}&
-\frac{1}{4}\sum_{n,n^\prime=-\infty}^\infty 
\bigg[-nn^\prime(a_n^0a_{n^\prime}^0+a_n^1a_{n^\prime}^1)\Big(\frac{2\pi}{T}\Big)^2+2\I E\frac{2\pi}{T}n\,(a_n^0a_{n^\prime}^1-a_{n^\prime}^0a_n^1)\bigg]\int_0^T\D\tau\;\E^{\I(n+n^\prime)\frac{2\pi}{T}\tau}
=\\={}&
-\frac{T}{4}\sum_{n,n^\prime=-\infty}^\infty\bigg[-nn^\prime(a_n^0a_{n^\prime}^0+a_n^1a_{n^\prime}^1)\Big(\frac{2\pi}{T}\Big)^2+2\I E\frac{2\pi}{T}n\,(a_n^0a_{n^\prime}^1-a_{n^\prime}^0a_n^1)\bigg]\delta_{n,-n^\prime}
=\\={}&
-\frac{T}{4}\sum_{n=-\infty}^\infty\bigg[n^2(|a_n^0|^2+|a_n^1|^2)\Big(\frac{2\pi}{T}\Big)^2
+2\I E\frac{2\pi}{T}n\,(a_n^0{a_n^1}^*-{a_n^0}^*a_n^1)\bigg]
=\\={}&
-\frac{T}{2}\sum_{n=1}^\infty\bigg[n^2(|a_n^0|^2+|a_n^1|^2)\Big(\frac{2\pi}{T}\Big)^2
+2\I E\frac{2\pi}{T}n\,(a_n^0{a_n^1}^*-{a_n^0}^*a_n^1)\bigg] .
\end{align}
Carrying out the path integral yields
\begin{align}
&\bigg(\mathcal{N}\int_\mathrm{P}[\D a][\D a^*]\E^{-\frac{T}{2}\sum_{n=1}^\infty[n^2(|a_n^0|^2+|a_n^1|^2)(\frac{2\pi}{T})^2+2\I E\frac{2\pi}{T}n\,(a_n^0{a_n^1}^*-{a_n^0}^*a_n^1)]}\bigg)^{-1}
=\\=&{}
\prod_{n=1}^\infty\frac{\left|\!\left|\begin{array}{cccc}
0&n^2(\frac{2\pi}{T})^2&0&2\I E\frac{2\pi}{T}n\\
n^2(\frac{2\pi}{T})^2&0&-2\I E\frac{2\pi}{T}n&0\\
0&-2\I E\frac{2\pi}{T}n&0&n^2(\frac{2\pi}{T})^2\\
2\I E\frac{2\pi}{T}n&0&n^2(\frac{2\pi}{T})^2&0
\end{array}\right|\!\right|^\frac{1}{2}}{[n^2(\frac{2\pi}{T})^2]^2}
\label{eq:det}
=\\=&{}
\prod_{n=1}^\infty\frac{[n^2(\frac{2\pi}{T})^2]^2-(2E\frac{2\pi}{T}n)^2}{[n^2(\frac{2\pi}{T})^2]^2}
=
\prod_{n=1}^\infty\bigg[1-\frac{(ET)^2}{(n\pi)^2}\bigg]
=
\frac{\sin(ET)}{ET}.
\label{eq:prod}
\end{align}
\end{widetext}
Putting this back into the effective action we obtain
\be
(4\pi)^2\frac{w}{\mathcal{V}}
\rightarrow
\int_\varepsilon^\infty\frac{\D T}{T^3}\;\E^{-m^2T}\frac{ET}{\sin(ET)} ,
\label{eq:effactE}
\ee
where $\mathcal{V}$ stands for the volume $\int\D^4x_0$, which factors out, as the integrand is translationally invariant and thus does not depend on $x_0$. (The most important difference for fermions is that the sin function would be replaced by a tan.) The vacuum persistence amplitude is given by twice the imaginary part of the effective action. The imaginary part arises from the poles of the integrand away from $T=0$. The $T=0$ pole is removed by renormalisation, which does not influence the imaginary part, that signals the decay of the vacuum. For a magnetic field, the entries on the diagonal from the southwest to the northeast of the matrix in (\ref{eq:det}) would be real, the relative sign in the last infinite product would change, and the infinite product would evaluate to a hyperbolic sine instead of a trigonometric. Thus, there would be no poles and consequently no imaginary part. In the presence of an electric field the imaginary part reads
\be
\mathrm{Im}\int_\varepsilon^\infty\frac{\D T}{T^3}\;\E^{-m^2T}\frac{ET}{\sin(ET)}
=
2\pi E^2\sum_{l=1}^\infty\frac{(-)^{l+1}}{(l\pi)^2}\E^{-\frac{m^2}{|E|}l\pi} .
\label{eq:exact}
\ee
(For fermions there is no alternating sign, because of the tan in the place of the sin.) Contributions from higher poles are suppressed exponentially and by another factor of $l^{-2}$. The latter form of suppression persists even in the massless case. 

{The above result holds for an arbitrarily strong field but weak coupling $e$ (which, for the sake of brevity, we have absorbed in the electric field $E$). At weak fields the result is dominated by the $l=1$ term. In this case going to larger couplings, according to \cite{Affleck:1982}, requires including all Coulomb exchanges, which leads to an additional factor of $\E^{e^2/4}$.}

In this particular setup the exponential tunneling factor is accessible via a short cut \cite{Affleck:1982,Dunne:2005sx}: Consider the worldline action including the mass term and carry out the $T$ integration in the saddle-point approximation, 
\be
T\rightarrow\frac{1}{2m}\sqrt{\int_0^1\D\hat\tau\;\dot y^2},
\label{eq:saddleT}
\ee
ignoring the negative powers of $T$ in the integration measure,
\begin{align}\nonumber
&-\int_0^T\D\tau\;\Big[\frac{\dot y^2}{4}+m^2-\frac{E}{2}(\dot y^0y^1-y^0\dot y^1)\Big]
=\\={}&
\label{eq:Tdep}
-\int_0^1\D\hat\tau\;\Big[\frac{\dot y^2}{4T}+m^2T-\frac{E}{2}(\dot y^0y^1-y^0\dot y^1)\Big]
\stackrel{(\ref{eq:saddleT})}{\rightarrow}\\\stackrel{(\ref{eq:saddleT})}{\rightarrow}{}&
-m\sqrt{\int_0^1\D\hat\tau\;\dot y^2}
+
\frac{E}{2}\int_0^1\D\hat\tau\;(\dot y^0y^1-y^0\dot y^1).
\label{eq:afterT}
\end{align}
Next, adopt a so-called worldline-instanton ansatz
\be
[y^0,y^1,y^2,y^3]=[R\sin(2\pi l\hat\tau),R\cos(2\pi l\hat\tau),0,0],
\label{eq:wlansatz}
\ee
$l\in\mathbbm{N}$, and put this ansatz into the worldline action on the $T$ saddle point (\ref{eq:afterT}),
\begin{align}\nonumber
&-m\sqrt{\int_0^1\D\hat\tau\;\dot y^2}
+
\frac{E}{2}\int_0^1\D\hat\tau\;(\dot y^0y^1-y^0\dot y^1)
\stackrel{(\ref{eq:wlansatz})}{\rightarrow}\\\stackrel{(\ref{eq:wlansatz})}{\rightarrow}{}&
-2\pi l\Big(mR-\frac{E}{2}R^2\Big)
\stackrel{(\ref{eq:saddleE})}{\rightarrow}
-\pi l\frac{m^2}{E},
\label{eq:onsaddle}
\end{align}
where in the last step we have put $R$ to its value on the saddle point,
\be
R\rightarrow\frac{m}{E}.
\label{eq:saddleE}
\ee
Hence, we have obtained the same exponent in two different ways. Here it was obtained by two consecutive saddle-point approximations. In the above exact computation the exponent arose by evaluating the mass dependent exponential at the poles of the rest of the integrand; those poles, in turn, appeared where the fluctuation determinant vanished, i.e., where the integral did not have a Gaussian suppression. Furthermore, in the exact calculation, we first performed the path integral and then the $T$ integration; in the second, we first approximated the $T$ integration and then replaced the functional integration by adjusting particular test functions. {(The use of the concept of worldline instantons does not depend on sticking to the latter order of integrations, though \cite{Dunne:2005sx}.)} In (\ref{eq:Tdep}) only the mass and kinetic terms were $T$ dependent. 
{Generally, the saddle-point approximation for the $T$ integration amounts to a weak-field approximation. Nevertheless, it reproduces exactly the exponents from the unapproximated $T$ integration (\ref{eq:exact}). Moreover, the result is accurate surprisingly far outside the weak-field regime \cite{Dunne:2005sx}, i.e., also for $E=O(m^2)$.}

Naturally $E$ can have either sign. In order to have positive saddle-point values of $R$, $E$ has to be positive. For negative $E$ we would have to choose a different worldline-instanton ansatz where the path is oriented in the opposite sense, i.e., with sine and cosine interchanged or with negative sign for the arguments. Then the relative sign of the terms in the rounded brackets of (\ref{eq:onsaddle}) would be + and the saddle-point value of $R$ would be $-\frac m E$. Thus, combining all cases we would have $R\rightarrow\frac{m}{|E|}$ and the final exponent $-\pi l\frac{m^2}{|E|}$.
This assessment is fully consistent with the first derivation, where already (\ref{eq:onsaddle}) only depends on $|E|$ and only poles at positive values of $T$ contribute due to the range of integration.
Furthermore, the first computation probed both orientations of the path simultaneously and the result decomposed into a product of the computations for one value of $n$ at a time, which is why an ansatz (\ref{eq:wlansatz}) where the single frequencies are treated separately can be considerd. As both orientations are treated at once, the individual factors in (\ref{eq:prod}) have two zeros at $T=\pm\pi n/E$ and the sign has to be chosen where $T$ is positive, i.e., $T=\pi n/|E|$. 

\section{Holography without strings attached\label{sec:nostrings}}

In this section we give a brief outline of our description of hadrons on the worldline and refer the reader to \cite{Dietrich:2014...} for more details. The basic observation behind this description of the phenomenology of strongly interacting gauge theories at low energies is that diagrams with the lowest possible number of transverse gluons are the dominant ones. In hadron-hadron scattering processes, for example, the exchange and annihilation of quarks are found to dominate over the exchange of gluons \cite{White:1994tj}. The opposite assumption, i.e., that gluons dominate, led to the Landshoff paradox \cite{Landshoff:1974ew}, which was resolved by recognising that they actually did not \cite{Gunion:1972qi}. Furthermore, the Okubo-Zweig-Iizuka rule \cite{Okubo:1963fa} states that the contributions to scattering processes of diagrams that fall apart when all gluon lines are removed is suppressed. These phenomena could be explained by the conditional survival of a perturbative counting scheme at low energies, for which independent indications also exist \cite{Dokshitzer:1998qp}. Finally, this is also consistent with the fact that hadrons can be characterised by their valence quark content and that we do not observe a plethora of multiquark states, hybrids, or glueballs. Also excitation spectra of quarkonium and positronium are surprisingly similar despite their vastly different energy scales.
  
\begin{figure}[h]
\centerline{%
\includegraphics[width=\columnwidth]{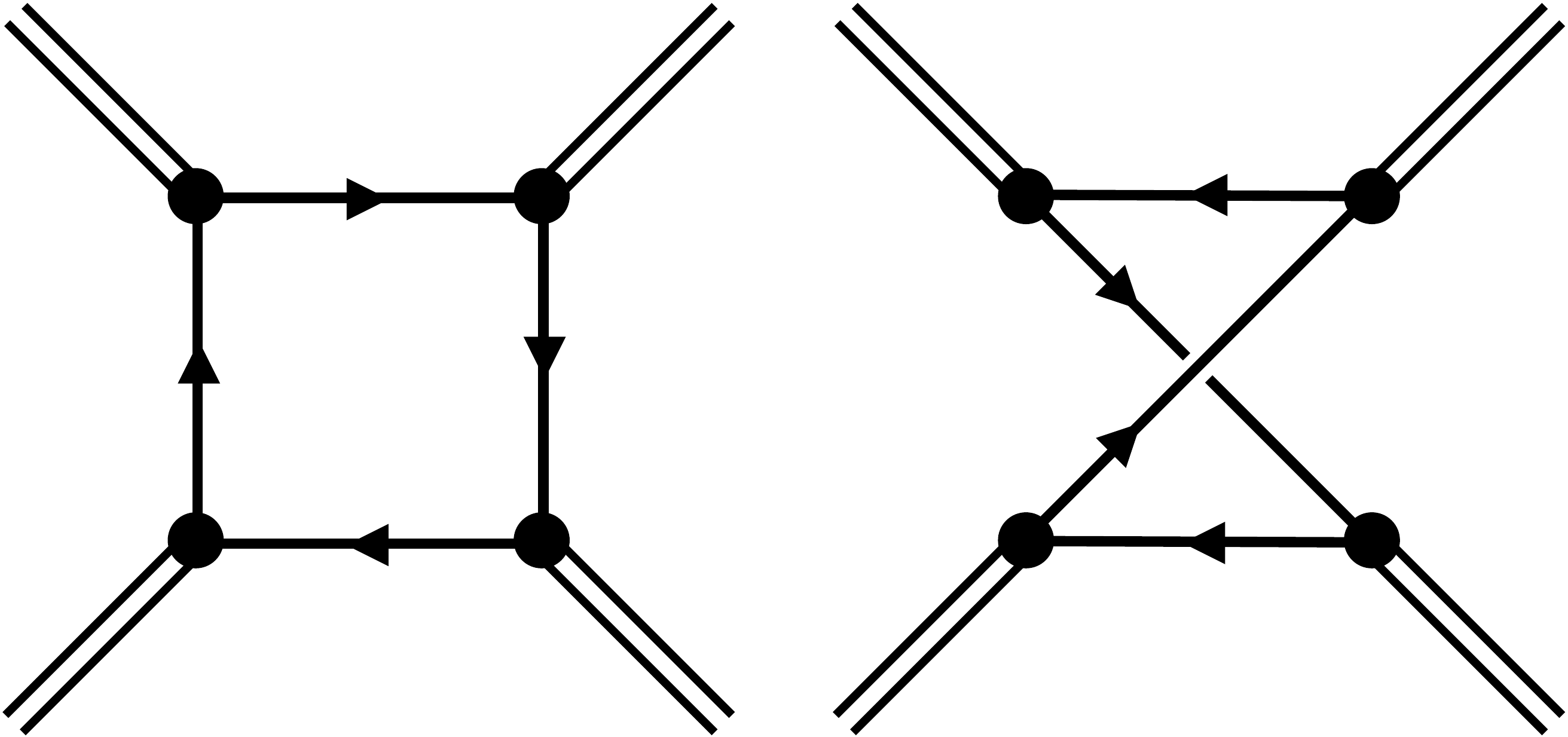}}
\caption{Some dominant diagrams. Double lines are for hadrons, single lines for quarks. (Taken from \cite{Dietrich:2014...}.)}
\label{fig}
\end{figure}

For vector mesons the dominant diagrams, like those shown in Fig.~\ref{fig},
are contained in the effective action (\ref{eq:ads?}) after replacing the gauge field $A^\mu$ by a vector source $V^\mu$. (It is straightforward to include also other mesons, i.e., scalars etc., which we refrain from here for the sake of simplicity.) Already at this point (\ref{eq:ads?}) possesses many parallels to the AdS/QCD \cite{Polchinski:2000uf,Karch:2006pv} description of the hadron spectrum: It takes the form of a Lagrangian density integrated over an AdS$_5$ space
\begin{align}
w&=\int \D^5x \sqrt{g}\,\E^{-m^2T}\mathcal{L},
\\
\Lag
&=
\frac{\mathcal{N}}{(4\pi)^2}
\int_\mathrm{P}[\D y]\;\E^{-\int_0^T\D\tau[\frac{\dot y^2}{4}+\I \dot y \cdot V(x+y)]} ,
\label{eq:lag}
\end{align}
with an additional warp factor $\E^{-m^2T}$. Here
\be
\D s^2\overset{g}{=}-\frac{\D T^2}{4T^2}+\frac{\D x\cdot\D x}{T}.
\label{eq:patch}
\ee
The interaction part is a Wilson line, which is invariant under local transformations $V^\mu\rightarrow\Omega[V^\mu+\I\Omega^\dagger(\partial^\mu\Omega)]\Omega^\dagger$; i.e., hidden local symmetry \cite{Bando:1984ej} emerges.
Furthermore, to leading nontrivial order in the so-called inverse mass expansion \cite{Schubert:2001he},
\be
w_{I\!I}
=
\frac{-1}{6(4\pi)^2}\:\mathrm{tr}\int\D^5x\sqrt{g}\:\E^{-m^2T}g^{\mu\kappa}g^{\nu\lambda}V_{\mu\nu}V_{\kappa\lambda} ,
\label{eq:T2}
\ee
which coincides with the corresponding expression in soft-wall AdS/QCD \cite{Karch:2006pv} for a flat profile function, i.e., for $\tilde v(p,T)\equiv 1$, in $\tilde{\mathcal{V}}^\mu(p,T)=\tilde V^\mu(p)\tilde v(p,T)$, where $\tilde{\mathcal{V}}^\mu(p,T)$ is the bulk vector field in mixed Fourier representation. (A numeric prefactor can be absorbed into the normalisation of the fields or a coupling constant, and the scales of conformal-symmetry breaking must be identified.) $V_{\mu\nu}$ stands for the field-strength tensor constructed from $V_\mu$ and $g^{\mu\nu}$ for the inverse AdS$_5$ metric belonging to (\ref{eq:patch}).
Even the divergence structure at small Schwinger proper time $T=\varepsilon$ coincides with that of AdS/QCD if $\varepsilon$ is the position of the UV brane.

Bottom-up AdS/QCD models describe the hadron spectrum surprisingly well \cite{Polchinski:2000uf,Karch:2006pv,Da Rold:2005zs}. They are inspired by the Maldacena conjecture \cite{Maldacena:1997re} and related exact dualities. None of these examples, however, has the particle content of QCD. Therefore, one is forced to rely on bottom-up models, which are not derived from first principles, and the role played by their ingredients, e.g., their extra dimension, remains to be clarified, which was a major motivation for our study. Of course AdS$_5$ has the same isometries as the conformal group over four-dimensional Minkowski space. As a consequence, these models have conformal symmetry as a first approximation, which is shared by massless classical QCD. Otherwise conformality is anomalous in QCD. (For quasiconformal technicolor models \cite{walk} it is an even better approximation and has been used in this context \cite{Hong:2006si}.)

In its present form the worldline effective action shows a threshold behaviour for virtualities above $4m^2$, while the soft-wall AdS/QCD expression exhibits a tower of states.
In \cite{Dietrich:2014...} we studied how to obtain a bound-state spectrum in the worldline approach. To this end, as a first experiment, we introduced artificially a tower of states by a mere change of variables, which, by definition, may not influence the physical result, and traced its effect through the effective action all the way to the worldline action (the exponent). The substitution induces a repulsive harmonic oscillator, which cancels exactly the effect of the artificially included power of states. In order to have a physical effect, the coefficients in the tower and the oscillator must be detuned. This includes the cases in which one of the parameters is zero, e.g., where there is only a harmonic oscillator and no tower of states. In particular the above indicates that the tower of states is linked to a harmonic oscillator term $\propto\int_0^T\D\tau\,y^2$ in the worldline action. (We will come back to this point in Sect.~\ref{sec:var}.)
The interpretation as mutual compensation between the tower of states and the repulsive harmonic oscillator is confirmed by noticing that the change of variables is a special case of transformations of conformal field theories that introduce a scale into the Lagrangian without affecting conformality, which is saved by a simultaneous change of the time variable \cite{de Alfaro:1976je}. [For another use of these transformations in light-front holography \cite{deTeramond:2008ht} see \cite{Brodsky:2013ar}. {Light-front holography identifies the extra dimension with $\zeta^2=x(1-x)\mathbf{b}_\perp^2$, where $x$ stands for the light-front momentum fraction of one of the constituents of the meson and $\mathbf{b}_\perp$ for the transverse separation of the constituents.}] (Coincidentally, the corresponding transformation of the source in \cite{Dietrich:2014...} is also known from computations in soft-wall AdS/QCD \cite{Karch:2006pv,Dietrich:2008ni}, where it moves the warp factor away from the kinetic term.) 

A harmonic oscillator term can also be seen as a two-body interaction,
\be
\int_0^1\D\hat\tau_1\D\hat\tau_2[y(\tau_1)-y(\tau_2)]^2
\stackrel{(\ref{eq:com})}{=}
2\int_0^1\D\hat\tau\, y(\tau)^2.
\label{eq:2body}
\ee
We can arrive at a similar interaction also in a different way \cite{Dietrich:2014...} and in particular with the tower of states as an emergent phenomenon \cite{Dietrich:2012un}: The absence of transverse gluons does not forestall the presence of an instantaneous interaction. In fact, omitting them initially amounts only to a poor lowest order approximation, as charged particles are always accompanied by the field they induce. Perturbation theory with bare particles leads to infrared divergences, which must be cured by including the Sudakov form factors, which are the manifestation of the Coulomb field of the scattering particle \cite{Collins:1989bt}. Analogously, a leading contribution to gauge-boson production in high-energy collisions is given by Weizs\"acker-Williams \cite{von Weizsacker:1934sx} radiation, which corresponds to their liberation from the Li\'enard-Wiechert \cite{Lienard:1898} fields of the charged particles.

Apart from the inverse-distance Coulomb piece an instantaneous linear piece aligned with the two constituents (here of a meson for the sake of concreteness) at $x_1^\mu$ and $x^\mu_2$, $A^0(x_1,x_2)=\frac{1}{2}\Lambda\delta(x^0_1-x^0_2)|\mathbf{x}_1-\mathbf{x}_2|$, is the only remaining additional term preserving full Poincar\'e invariance \cite{Dietrich:2012un}, which in this context automatically entails the stationarity of the action. This component corresponds to a nontrivial boundary condition at infinity, $\lim_{r\rightarrow\infty}F_{\mu\nu}^2=\Lambda^2$,
and is of the lowest order in the gauge coupling constant of all possible contributions. (This point, together with our entire approach as laid out at the beginning of the present section and in \cite{Dietrich:2014...}, is akin to the condensate picture of \cite{Shifman:1978bx}.) We can incorporate this field in the effective action (\ref{eq:ads?}) by averaging over the corresponding component of the gauge field,
\begin{align}
&\mathcal{N}_A\langle\E^{-\I\int_0^T\D\tau\,\dot y \cdot A}\rangle
=\\=
{}&\mathcal{N}_A\int[\D A]\,\E^{-\frac{1}{2}\int\D^4x\,A\cdot\Gamma^{-1}\cdot A}\E^{-\I\int_0^T\D\tau\,\dot y \cdot A}
=\\=
{}&\E^{-\frac{1}{2}\int_0^T\D\tau_1\D\tau_2\,\dot y_1\cdot\Gamma(y_1-y_2)\cdot\dot y_2}
\stackrel{[26]}{\rightarrow}\\\stackrel{[26]}{\rightarrow}
{}&\E^{\frac{\Lambda}{2}\int_0^T\D\tau_1\D\tau_2\,\delta(y_1^0-y_2^0)\dot y_1^0|\mathbf{y}_1-\mathbf{y}_2|\dot y_2^0},
\label{eq:delta}
\end{align}
where $\Gamma$ stands for the propagator, and the normalisation $\mathcal{N}_A=\langle 1\rangle^{-1}$ cancels the path integral for $\dot y^\mu\equiv0$. For the lowest Fock state, i.e., if the path only curves back once in the time ($y^0$) direction, integrating out one of the Schwinger parameters yields
\begin{align}
\mathcal{N}_A\langle\E^{-\I\int_0^T\D\tau\,\dot y \cdot A}\rangle
\supset
\E^{\frac{\Lambda}{2}\int_0^T\D\tau\,\mathrm{sgn}(\dot{\bar{y}}^0)\dot y^0|\mathbf{y}-\bar{\mathbf{y}}|}
=\E^{-\Lambda\times\mathrm{Area}},
\label{eq:area}
\end{align}
where $\mathbf{y}$ and $\bar{\mathbf{y}}$ denote the two points which at equal time $y^0$ are on opposite sides of the trajectory. The exponent is proportional to the absolute area enclosed by the trajectory. (An analogous result is obtained for a large number of colours in two dimensions \cite{'tHooft:1973jz}. For the oriented area and in two dimensions it could be linked to a static external magnetic field \cite{Cornwall:2003mt} and its linearly spaced Landau levels, which are again linked to a harmonic oscillator in a composite variable \cite{Greiner:1992bv}.) Together with the kinetic term for $y^0$ we can complete the square,
\begin{align}\nonumber
&(\dot y^0)^2-2\Lambda\:\mathrm{sgn}(\dot{\bar{y}}^0)\dot y^0|\mathbf{y}-\bar{\mathbf{y}}|
=\\=
{}&(\dot z^0)^2-\Lambda^2(\mathbf{y}-\bar{\mathbf{y}})^2
\label{eq:spatattrho}
=\\=
{}&(\dot z^0)^2-\Lambda^2({y}-\bar{{y}})^2 .
\label{eq:resemb}
\end{align}
Here $\dot z^0=\dot y^0-\Lambda\,\mathrm{sgn}(\dot{\bar{y}}^0)|\mathbf{y}-\bar{\mathbf{y}}|$ and $\dot{\bar{y}}^0=\partial_\tau y^0|_{\tau=\bar\tau}$ as well as by definition $y^0-\bar{y}^0\equiv0$. The last expression (\ref{eq:resemb}) bears strong resemblance with the two-body interpretation in (\ref{eq:2body}). [$\mathrm{sgn}(\dot{\bar{y}}^0)=-\mathrm{sgn}(\dot{{y}}^0)$ is always true, but for parametrisations where $\dot y^0=-\dot{\bar{y}}^0$ as well, which is a natural gauge choice in an equal-time description, $z^0$ is periodic and also $z^0-\bar z^0\equiv0$.]

Let us close this section with a few more remarks \cite{Dietrich:2014...}: In the present approach the holographic extension of the four-dimensional fields to the fifth dimension proceeds via a Wilson flow (gradient flow) \cite{Luscher:2009eq}, and if we ask for an optimal flow, e.g., in (\ref{eq:T2}), we reproduce exactly the AdS/QCD computation. ~ Repeating our worldline construction in the nonrelativistic case studied in the context of condensed matter physics \cite{Sachdev:2011wg} systematically reproduces the known extradimensional spacetime structure \cite{Son:2008ye}. ~ The worldline formalism can be gainfully analysed in the worldline-instanton framework, which we are going to discuss below. The latter framework can be related \cite{Dietrich:2007vw} to the Gutzwiller trace formula \cite{Gutzwiller:1971fy}, which describes quantum mechanical systems through classical attributes (generally approximately, but exactly for quadratic actions), i.e., periodic orbits, stability matrices, and Morse indices, analogously to the quantum mechanical approach to quantum field theory in the framework of holography.

\section{Worldline holographic Schwinger effect\label{sec:wh}}

In order to analyse the vacuum persistence in the presence of a confining interaction we include the interaction term from (\ref{eq:area}) in the action (\ref{eq:afterT}) on the saddle point of the $T$ integration. 
Subsequently putting in the worldline-instanton ansatz (\ref{eq:wlansatz}) yields for $l=1$ 
\begin{align}\nonumber
&-
m\sqrt{\int_0^1\D\hat\tau\;\dot y^2}
-
\frac{\Lambda}{2}\int_0^1\D\hat\tau\,|\dot y^0||\mathbf{y}-\bar{\mathbf{y}}|
+\\&+
\frac{E}{2}\int_0^1\D\hat\tau\;(\dot y^0y^1-y^0\dot y^1)
\label{eq:Fock1}
\stackrel{(\ref{eq:wlansatz})}{\rightarrow}\\\stackrel{(\ref{eq:wlansatz})}{\rightarrow}{}&\label{eq:beforeRsaddle}
-2\pi \big(mR-\sfrac{1}{2}|E|R^2\big)
-\Lambda\pi R^2
\stackrel{(\ref{eq:saddleRL})}{\rightarrow}\\\stackrel{(\ref{eq:saddleRL})}{\rightarrow}{}&
-\pi \frac{m^2}{|E|-\Lambda}\mathrm{~~~if~~~}|E|>\Lambda,
\label{eq:exponentL}
\end{align}
where the saddle-point value for $R$ is given by
\be
R\rightarrow\frac{m}{|E|-\Lambda}\mathrm{~~~if~~~}|E|>\Lambda .
\label{eq:saddleRL}
\ee
We have expressed the result in terms of $|E|$ bearing in mind that, as already discussed above, the orientation of the orbit has to be adjusted such that the saddle-point values of the integration variables stay within their range of integration, i.e., positive. In any case, i.e., trying both orientations, positive values for $R$ can only be achieved for $|E|>\Lambda$. Otherwise there is no imaginary part and the vacuum is stable. In the limit $\Lambda\rightarrow 0$ the original result is recovered.

{At the threshold, $|E|=\Lambda$, the worldline action on the saddle point diverges like $(|E|-\Lambda)^{-1}$. A divergent behaviour for the action at this point is also found in the holographic study \cite{Kawai:2013xya}. There, divergent terms $\propto(|E|-\Lambda)^{-2}$ as well as $\propto(|E|-\Lambda)^{-1}$ are found, and a numerical analysis indicates that the prefactor of the former term vanishes, if $\Lambda$ is not small against $m^2$.}

{Analogously to the derivation of (\ref{eq:onsaddle}) the saddle-point approximation leading to (\ref{eq:saddleRL}) is a priori a weak-field approximation, which here means that $|E|-\Lambda$ (if it is $>0$) has to be compared to $m^2$. This follows from a comparison of the combination of parameters in (\ref{eq:beforeRsaddle}) and (\ref{eq:onsaddle}). Thus especially the threshold behaviour is diagnosed accurately. Additionally, we saw that for (\ref{eq:saddleRL}) the exact computation gave the identical result for the exponents (and will do so again in Section \ref{sec:var}). This is very similar here, as we shall discuss before turning to Section \ref{sec:var}. Moreover, this particular saddle-point computation generally seems to have an extended range of validity \cite{Dunne:2005sx}.}

If we put the ansatz (\ref{eq:wlansatz}) for $l>1$ into (\ref{eq:Fock1}), we would get (\ref{eq:beforeRsaddle}) multiplied by an overall $l$, in the first two addends due to the derivatives and in the $\Lambda$ term because the area is covered $l$ times. This would lead to the same saddle-point condition (\ref{eq:saddleRL}) and to an overall factor of $l$ multiplying (\ref{eq:exponentL}). This, however, is not in accordance with (\ref{eq:delta}), as (\ref{eq:area}) was derived for only a single recurrence in the time direction. If we account properly for multiple recurrences, there appears an additional factor of $l$ in the $\Lambda$ term, accounting for the $l$ different contributions from evaluating the $\delta$ distribution in (\ref{eq:delta}). Taking stock for $l>1$ (\ref{eq:beforeRsaddle}) is multiplied by $l$ and additionally $\Lambda$ multiplied by $l$, which leads to the modified saddle point 
\be
R\rightarrow\frac{m}{|E|-l\Lambda}\mathrm{~~~if~~~}|E|>l\Lambda
\ee
and the exponent
$-\pi l\frac{m^2}{|E|-l\Lambda}$. This implies that for larger $l$ the condition for vacuum instability becomes harder and harder to meet and the sum over $l$ will cease before $|E|<l\Lambda$. 

This result can be cross-checked by putting a correspondingly constrained version of the parametrisation (\ref{eq:Fourier}),
\be
y^{0,1}
=
a_{+n}^{0,1}\;\E^{+\I n\frac{2\pi}{T}\tau}+a_{-n}^{0,1}\;\E^{-\I n\frac{2\pi}{T}\tau} ,
\label{eq:constrained}
\ee
(no sum over $n$) into the worldline action (\ref{eq:delta}). After exploiting rotational invariance to align the principal axes of the elliptic orbit with the coordinate axes ($a_n^0=a^0$ and $a_n^1=\I a^1$, where $a^0,a^1\in\mathbbm{R}$) this yields
\begin{align}
&-
\frac{1}{4}\int_0^T\D\tau\;[\dot y^2-2E(\dot y^0y^1-y^0\dot y^1)]
\nonumber+\\&+\nonumber
\frac{\Lambda}{2}\int_0^T\D\tau_1\D\tau_2\,\delta(y_1^0-y_2^0)\dot y_1^0|\mathbf{y}_1-\mathbf{y}_2|\dot y_2^0
=\\={}&
-n^2\pi|a|^2\frac{2\pi}{T}
+
4 E\pi n\,a^0a^1
-
4\Lambda n^2\pi|a^0a^1|
\label{eq:PHreload}
=\\={}&
-2\pi n|a|^2
\Big[
n\frac{\pi}{T}
-
E\,\sin(2\phi)
+
\Lambda n|\sin(2\phi)|
\Big]
\breve\rightarrow\\\breve\rightarrow{}&
-2\pi n|a|^2
\Big[
n\frac{\pi}{T}
-
|E|
+
\Lambda n
\Big]\stackrel{!}{=}0
\label{eq:sadpo}
\\{}&\Leftrightarrow
T=\frac{n\pi}{|E|-n\Lambda}
~~~\Rightarrow~~~\E^{-m^2T}\rightarrow\E^{-n\pi\frac{m^2}{|E|-n\Lambda}},
\label{eq:expex}
\end{align}
where we transformed to polar coordinates $a^0=|a|\cos\phi$ and $a^1=|a|\sin\phi$. 
The sum of the last two $\phi$ dependent addends in the square brackets can only become negative if $|E|>n\Lambda$. Only then can the entire expression become zero for positive values of $T$, which corresponds to poles in the integrand of the $T$ integration and leads to an imaginary part for the effective action. The exact exponent (\ref{eq:expex}) is reproduced at the saddle point of the $\phi$ integration (\ref{eq:sadpo}). (The rest of the $\phi$ dependence contributes merely to the fluctuation prefactor.)

{In Sections \ref{sec:persist} and \ref{sec:var} the saddle-point approximation yields exactly the same exponent as the full computation, because there we are dealing with quadratic Lagrangians. Here, this is not the case for a general field $y^\mu$ as given by (\ref{eq:Fourier}). 
Nevertheless, when adding a perturbation $\delta$ with $n^\prime\neq n$ to (\ref{eq:constrained}) we do not get any terms nondiagonal in $n$ from the kinetic and electric field terms as before, and the the contributions to the potential term are
$O(\delta^2)$ and $O(\frac{\delta^4}{|a|^2})$.
The imaginary part for the configurations (\ref{eq:constrained}) arose where the fluctuation determinant vanishes,i.e., where the typical value of $|a|^2$ that contributes to the path integral has no Gaussian cutoff and is generally large. Thus, the $O(\frac{\delta^4}{|a|^2})$ term is small for finite $\delta$ as are all terms with even higher powers of $\frac{\delta^2}{|a|^2}$. 
Hence, near the poles, we again effectively have a quadratic Lagrangian for which the saddle-point approximation is exact, and the position of the poles and thus the exponents of the imaginary part are captured well in this framework. (The above does not say that the approximation should be as good for the real part, since it gets contributions from everywhere, especially from away from the poles, but we are only concerned with the imaginary part here.)  The Gaussian cutoff for large values of $\delta^2$ will not be absent simultaneously to that of $|a|^2$, as $n^\prime\neq n$ by definition. This is why $\frac{\delta^2}{|a|^2}$ is very small. Reciprocally, where the Gaussian suppression for $n^\prime$ goes away there will be a Gaussian suppression for $n$, and we reproduce the pole for $n^\prime$. Consequently, the different values of $n$ effectively do not mix. The final $T$ integration is exact and yields the sum over the various contributions.}

Summarising, due to the presence of the confining interaction, the vacuum is only unstable for electric fields above a threshold given by the strength of the aforesaid interaction. This is a behaviour that is also present in analogous studies directly based on AdS/CFT holography {\cite{Sato:2013dwa,Kawai:2013xya,Hashimoto:2013mua}}. 

\subsection{Variations\label{sec:var}}

Let us compare the result to the case of an attractive (with respect to the classical equations of motion for $y^\mu$) harmonic oscillator {discussed at the beginning of Sect.~\ref{sec:nostrings} (page 4, second column),}
\be
\Lag
\rightarrow
\frac{\mathcal{N}_c}{(4\pi)^2}
\int_\mathrm{P}[\D y]\;\E^{-\int_0^T\D\tau[\frac{\dot y^2}{4}-\frac{c^2}{4}y^2-\frac{E}{2}(\dot y^0y^1-y^0\dot y^1)]} ,
\label{eq:lagc}
\ee
where the normalisation {$\mathcal{N}_c$} cancels the path integral for $E=0$. [We start by looking at an attractive sign also because this is what arises by the completion of the square (\ref{eq:spatattrho}) in the approach studied at the beginning of this section.] Using the parametrisation (\ref{eq:Fourier}) the exponent becomes
\begin{widetext}
\begin{align}\nonumber
&-\frac{1}{4}\int_0^T\D\tau\;\Big[\dot y^2-\frac{c^2}{4}y^2-2E(\dot y^0y^1-y^0\dot y^1)\Big]
=\\={}&
-\frac{1}{4}\sum_{n,n^\prime=-\infty}^\infty a_n\cdot a_{n^\prime}\bigg\{\bigg[-nn^\prime\Big(\frac{2\pi}{T}\Big)^2-\frac{c^2}{4}\bigg]+2\I E\frac{2\pi}{T}n\,(a_n^0a_{n^\prime}^1-a_{n^\prime}^0a_n^1)\bigg\}\int_0^T\D\tau\;\E^{\I(n+n^\prime)\frac{2\pi}{T}\tau}
=\\={}&
-\frac{T}{4}\sum_{n,n^\prime=-\infty}^\infty a_n\cdot a_{n^\prime}\bigg\{\bigg[-nn^\prime\Big(\frac{2\pi}{T}\Big)^2-\frac{c^2}{4}\bigg]+2\I E\frac{2\pi}{T}n\,(a_n^0a_{n^\prime}^1-a_{n^\prime}^0a_n^1)\bigg\}\delta_{n,-n^\prime}
=\\={}&
-\frac{T}{4}\sum_{n=-\infty}^\infty\bigg\{|a_n|^2\bigg[n^2\Big(\frac{2\pi}{T}\Big)^2-\frac{c^2}{4}\bigg]
+2\I E\frac{2\pi}{T}n\,(a_n^0{a_n^1}^*-{a_n^0}^*a_n^1)\bigg\}
=\\={}&
-\frac{T}{2}\sum_{n=1}^\infty\bigg\{|a_n|^2\bigg[n^2\Big(\frac{2\pi}{T}\Big)^2-\frac{c^2}{4}\bigg]
+2\I E\frac{2\pi}{T}n\,(a_n^0{a_n^1}^*-{a_n^0}^*a_n^1)\bigg\} .
\label{eq:atho}
\end{align}
Carrying out the path integral (\ref{eq:pint}) we find
\begin{align}
&\bigg(\mathcal{N}_c\int_\mathrm{P}[\D a][\D a^*]\E^{-\frac{T}{2}\sum_{n=1}^\infty\{|a_n|^2[n^2(\frac{2\pi}{T})^2-\frac{c^2}{4}]+2\I E\frac{2\pi}{T}n\,(a_n^0{a_n^1}^*-{a_n^0}^*a_n^1)\}}\bigg)^{-1}
=\\={}&
\prod_{n=1}^\infty\frac{\left|\!\left|\begin{array}{cccc}
0&n^2(\frac{2\pi}{T})^2-\frac{c^2}{4}&0&2\I E\frac{2\pi}{T}n\\
n^2(\frac{2\pi}{T})^2-\frac{c^2}{4}&0&-2\I E\frac{2\pi}{T}n&0\\
0&-2\I E\frac{2\pi}{T}n&0&n^2(\frac{2\pi}{T})^2-\frac{c^2}{4}\\
2\I E\frac{2\pi}{T}n&0&n^2(\frac{2\pi}{T})^2-\frac{c^2}{4}&0
\end{array}\right|\!\right|^\frac{1}{2}}{[n^2(\frac{2\pi}{T})^2-\frac{c^2}{4}]^2}
=\\={}&
\prod_{n=1}^\infty\frac{[n^2(\frac{2\pi}{T})^2-\frac{c^2}{4}]^2-(2E\frac{2\pi}{T}n)^2}{[n^2(\frac{2\pi}{T})^2-\frac{c^2}{4}]^2}
=
\prod_{n=1}^\infty\bigg\{1-\frac{(2E\frac{2\pi}{T}n)^2}{[n^2(\frac{2\pi}{T})^2-\frac{c^2}{4}]^2}\bigg\}
=\\={}&
\frac{\sin(\frac{T}{4}\sqrt{c^2+4E^2}+\frac{ET}{2})\sin(\frac{T}{4}\sqrt{c^2+4E^2}-\frac{ET}{2})}{\sin^2(\frac{cT}{4})}.
\label{eq:attr}
\end{align}
\end{widetext}
As we could already expect from the sign in front of $\frac{c^2}{4}$ everywhere above, if we compare it to the sign of the $\Lambda$ dependent term in (\ref{eq:PHreload}) there is no threshold behaviour in this setup. The sine functions in the numerator have zeros for positive values of $T$, which in the final expression will lead to an imaginary part for the effective action. The magnitude of the effect, however, is difficult to estimate, e.g., from the first few poles. This is because, while the position of the poles is obvious, the prefactors of the exponential terms also depend on the value of the other sine at the position of the pole, and this can compensate the stronger exponential suppression for large $n$ if two zeros of the two sine functions happen to be close to each other. On top of that, the product of the two sine functions leads to a beat which is an additional source for changing signs for the contributions from the various poles. We will revisit this point in Sect.~\ref{sec:a0neq0}.

At variance, for a repulsive harmonic oscillator
\be
\Lag
\rightarrow
\frac{\mathcal{N}_{\I c}}{(4\pi)^2}
\int_\mathrm{P}[\D y]\;\E^{-\int_0^T\D\tau[\frac{\dot y^2}{4}+\frac{c^2}{4}y^2-\frac{E}{2}(\dot y^0y^1-y^0\dot y^1)]} ,
\label{eq:lagic}
\ee
instead of (\ref{eq:attr}) we would get 
\begin{align}
-\frac{\sin(\frac{T}{4}\sqrt{4E^2-c^2}+\frac{ET}{2})\sin(\frac{T}{4}\sqrt{4E^2-c^2}-\frac{ET}{2})}{\sinh^2(\frac{cT}{4})}
\nonumber=\\=
\frac{\cosh(\frac{T}{2}\sqrt{c^2-4E^2})-\cos(ET)}{2\sinh^2(\frac{cT}{4})}.
\end{align}
For $c^2>4E^2$ we can see from the second line that there are no zeros for positive $T$. Hence the effective action will not have an imaginary part. Consequently the vacuum is stable against particle production. For the opposite case $c^2\le 4E^2$ the presence of zeros for positive $T$ and therefore the instability of the vacuum can be read off from the first line of the previous expression. Hence, the phenomenology of this setup is in this respect close to the one studied at the beginning of section \ref{sec:nostrings}. The repulsive harmonic oscillator postpones the appearance of closed classical orbits to after a threshold value for the electric field and thus stabilises the vacuum as a confining interaction should.

\subsubsection{$a_0^\mu\neq0$\label{sec:a0neq0}}

In \cite{Dietrich:2014...} we had studied the response of the system to the artificial introduction of a tower of states by means of a change of variables, to find out which kind of worldline potential belongs to a linearly spaced tower. Concretely, 
\be
\label{eq:covT}
cT=\E^{c\Theta}-1
\ee
led to the desired
\be
\int_\varepsilon^\infty\frac{\D T}{T}f(T)
=
c\int_{\varepsilon}^\infty\D\Theta\frac{f[T(\Theta)]}{1-\E^{-c\Theta}},
\label{eq:tower}
\ee
which corresponds to a sum of linearly spaced states with squared masses being integer multiples of the parameter $c$. 
In order to have $\Theta$ as the upper limit of integration in the worldline action we repeated this substitution for the variable $\theta$,
\be
c\tau=\E^{c\theta}-1 ,
\label{eq:covtau}
\ee
which changed the integrand of the worldline action. A standard kinetic term was restored through 
\be
y^\mu=\E^{c\theta/2}\xi^\mu,
\ee
which resulted in
\be
\int_0^T\D\tau\Big(\frac{\D y}{\D\tau}\Big)^2
=
\int_0^\Theta\D\theta\Big[\Big(\frac{\D\xi}{\D\theta}\Big)^2+\frac{c^2}{4}\xi^2+c{\frac{\D(\xi^2)}{\D\theta}}\Big].
\label{eq:coco}
\ee
The total derivative gives rise to a surface term, which vanishes for the starting-point conventions $y^\mu(0)=0=y^\mu(T)$, but generally not for the centre-of-mass conventions used so far for analysing the harmonic oscillator term. We could now either keep the surface term and redo the analysis in the centre-of-mass conventions or study harmonic oscillators without the surface term, but for starting-point conventions. In the latter case $a_0^\mu=\sum_{n=1}^\infty(a_n^\mu+{a_n^\mu}^*)$ instead of $a_0^\mu=0$. $a_0^\mu$ contributes to the potential term, for instance, in (\ref{eq:atho}). We can express (\ref{eq:atho}) as $a^\top Ma$, where $a$ is the vector of the coefficients $a_n$ grouped as follows,
\be
a^\top=(\dots,{a_{n-1}^1}^*,a_n^0,{a_n^0}^*,a_n^1,{a_n^1}^*,a_{n+1}^0,\dots) .
\ee
$M$ can be decomposed into
\be
M=A+\frac{c^2T}{16}(uu^\top+vv^\top),
\ee
where in the vectors $u$ and $v$ the following patterns are repeated for every value of $n$,
\begin{align}
u^\top&=\frac{c}{4}\sqrt{T}(\dots,1,1,0,0,\dots),\\
v^\top&=\frac{c}{4}\sqrt{T}(\dots,0,0,1,1,\dots).
\end{align}
$A$ is block diagonal with $4\times4$ submatrices $A_n$ along the diagonal,
\begin{widetext}
\be
A_n=-\frac{T}{4}\left(\begin{array}{cccc}
0&n^2(\frac{2\pi}{T})^2-\frac{c^2}{4}&0&2\I E\frac{2\pi}{T}n\\
n^2(\frac{2\pi}{T})^2-\frac{c^2}{4}&0&-2\I E\frac{2\pi}{T}n&0\\
0&-2\I E\frac{2\pi}{T}n&0&n^2(\frac{2\pi}{T})^2-\frac{c^2}{4}\\
2\I E\frac{2\pi}{T}n&0&n^2(\frac{2\pi}{T})^2-\frac{c^2}{4}&0
\end{array}\right).
\ee
\end{widetext}
We have to compute
\be
||M||=\Big|\Big|A+\frac{c^2T}{16}(uu^\top+vv^\top)\Big|\Big|,
\ee
which according to the matrix determinant lemma and the Sherman-Morrison formula can be expressed as
\begin{align}
&||M||\big/||A||
=\\={}&\nonumber
[(1+u^\top A^{-1}u)(1+v^\top A^{-1}v)-(u^\top A^{-1}v)(v^\top A^{-1}u)].
\end{align}
The inverse of $A_n$ is given by
\be
{A_n}^{-1}
=
-\frac{4}{T}\Big\{\Big[\frac{c^2}{4}-n^2\Big(\frac{2\pi}{T}\Big)^2\Big]^2-\Big(2nE\frac{2\pi}{T}\Big)^2\Big\}^{-1}A_n.
\ee
Consequently,
\begin{align}
u^\top A^{-1}u&=\frac{c^2}{4}\sum_{n=1}^\infty\frac{2[\frac{c^2}{4}-n^2(\frac{2\pi}{T})^2]}{[\frac{c^2}{4}-n^2(\frac{2\pi}{T})^2]^2-(2nE\frac{2\pi}{T})^2}=\nonumber\\=v^\top A^{-1}v&
\end{align}
and
\be
u^\top A^{-1}v=0=v^\top A^{-1}u.
\ee
$||A||$ is already known from above,
\be
||A||=\prod_{n=1}^\infty\Big(\frac{T}{4}\Big)^4\Big\{\Big[\frac{c^2}{4}-n^2\Big(\frac{2\pi}{T}\Big)^2\Big]^2-\Big(2nE\frac{2\pi}{T}\Big)^2\Big\}^2.
\ee
Putting everything together and normalising with respect to the $E=0$ case yields
\begin{align}
&\sqrt{||M||\big/||M||_{E=0}}
=\\={}&
\frac
{1+\frac{c^2}{4}\sum_{n=1}^\infty\frac{2[\frac{c^2}{4}-n^2(\frac{2\pi}{T})^2]}{[\frac{c^2}{4}-n^2(\frac{2\pi}{T})^2]^2-(2nE\frac{2\pi}{T})^2}}
{1+\frac{c^2}{4}\sum_{n=1}^\infty\frac{2}{\frac{c^2}{4}-n^2(\frac{2\pi}{T})^2}}
\label{eq:unres}\\\nonumber{}&
\frac
{\prod_{n=1}^\infty\{[\frac{c^2}{4}-n^2(\frac{2\pi}{T})^2]^2-(2nE\frac{2\pi}{T})^2\}}
{\prod_{n=1}^\infty[\frac{c^2}{4}-n^2(\frac{2\pi}{T})^2]^2}
=\\={}&\label{eq:res}
\frac
{\frac{c^2T\sqrt{c^2+4E^2}}{8(c^2+4E^2)}\sum_\pm\pm\cot[(\frac{E}{2}\pm\frac{1}{4}\sqrt{c^2+4E^2})T]}
{\frac{cT}{4}\cot(\frac{cT}{4})}
\\&\nonumber
\frac{\sin[(\frac{1}{4}\sqrt{c^2+4E^2}+\frac{E}{2})T]\sin[(\frac{1}{4}\sqrt{c^2+4E^2}-\frac{E}{2})T]}{\sin^2(\frac{cT}{4})}
=\\={}&
\frac{c}{\sqrt{c^2+4E^2}}
\frac{\sin[\sfrac{1}{2}\sqrt{c^2+4E^2}T]}{\sin(\frac{cT}{2})} .
\label{eq:canceled}
\end{align}
As before, the zeros of this expression determine the values of the exponents of the tunneling factors. The second fraction in (\ref{eq:unres}) is the result for $a_0^\mu=0$. The zeros of its numerator, however, are now compensated by poles in the numerator of the first fraction. Hence, the zeros of the numerator of the first fraction will now determine the positions of the poles. We can carry out all the infinite sums and products exactly to find (\ref{eq:res}), which simplifies to (\ref{eq:canceled}). As a result we see that for the starting-point convention we have evenly spaced poles, as in (\ref{eq:prod}). The beat has been removed. The exponent is an integer multiple of $-\pi m^2/(E^2+\frac{c^2}{4})^{1/2}$. For this sign of the harmonic oscillator potential the decay of the vacuum is still not postponed to larger field strengths. To the contrary the strength of the harmonic oscillator makes the contributions arrive in faster succession. 
{For the opposite sign we find $-\pi m^2/(E^2-\frac{c^2}{4})^{1/2}$, i.e., a delay of the onset of the vacuum decay. The worldline action diverges again at the threshold; however not $\propto(|E|-\Lambda)^{-1}$ as above, but $\propto(E^2-\frac{c^2}{4})^{-1/2}$.}

\subsubsection{Generalisations}

For a magnetic field, $E^2\rightarrow-B^2$ in (\ref{eq:canceled}). The vacuum becomes stable for $4B^2>c^2$, because then the sine turns into a hyperbolic sine; but the vacuum is not stable for smaller values of $B^2$. (For $E^2=0=B^2$ there is no particle production, as then the determinant cancels exactly against the normalisation.) 

For the opposite sign of the oscillator potential, $c^2\rightarrow-c^2$, the vacuum is stable for magnetic fields of arbitrary strength, since the argument of the square root in the sine in (\ref{eq:canceled}) is always negative. 

In the setup of the beginning of Sect.~\ref{sec:wh} a magnetic field corresponds to the replacement $E\rightarrow\I B$, which moves the poles away from the real axis and leads to a vanishing imaginary part. Hence, also in this setup the vacuum is stable in the presence of a magnetic field.

In the simultaneous presence of an electric and a magnetic field with $\mathbf{E}\cdot\mathbf{B}\neq0$, there exists a frame in which $\mathbf{E}||\mathbf{B}$. Then we get the product of two times (\ref{eq:canceled}), once for the electric field and once with $E^2\rightarrow -B^2$. We can express the result in a frame independent manner by reconstructing $E$ and $B$ from the relativistic invariants $\mathbf{E}^2-\mathbf{B}^2$ and $\mathbf{E}\cdot\mathbf{B}$,
\begin{align}
2E^2\rightarrow\sqrt{(\mathbf{E}^2-\mathbf{B}^2)^2+4(\mathbf{E}\cdot\mathbf{B})^2}+(\mathbf{E}^2-\mathbf{B}^2),
\\
2B^2\rightarrow\sqrt{(\mathbf{E}^2-\mathbf{B}^2)^2+4(\mathbf{E}\cdot\mathbf{B})^2}-(\mathbf{E}^2-\mathbf{B}^2).
\end{align}

\section{Conclusion\label{sec:concl}}

In conclusion, our worldline-holographic model for hadrons predicts the delay to larger field strength of vacuum decay by pair production in the presence of an external field. This delay stems from the presence of a confining force. It leads to the suppression from contributions of spatially large particle loops. In the worldline-instanton picture it forestalls the existence of classical periodic orbits for insufficiently large external fields.
{This is in line with findings in analogous studies setting out directly from AdS/CFT holography {\cite{Sato:2013dwa,Kawai:2013xya,Hashimoto:2013mua}}.}

\section*{Acknowledgments}

D.D.D.~would like to thank
Stan Brodsky,
Guy de T\'eramond,
Luigi Del Debbio,
Gia Dvali,
C\'esar Gomez,
Alexander Gu{\ss}mann,
Stefan Hofmann,
Paul Hoyer,
Michael Kopp,
Matti J\"arvinen,
Florian Niedermann,
Joachim Reinhardt,
Tehseen Rug,
Andreas Sch\"afer,
Robert Schneider,
Karolina Socha,
Christian Weiss,
Nico Wintergerst,
and
Roman Zwicky 
for inspiring and informative discussions.
The work of D.D.D.~was supported by the Humboldt Foundation.

\end{document}